\documentclass[twocolumn,aps]{revtex4}

\usepackage{graphicx}
\usepackage{dcolumn}
\usepackage{bm}
\usepackage{times}

\begin{document}
\title{Monte-Carlo simulation of localization dynamics of excitons in ZnO and CdZnO quantum well structures}

\author{T. Makino$^{a)}$}
\email[E-mail: ]{makino@sci.u-hyogo.ac.jp}
\affiliation{Department of Material Science, University of Hyogo, Kamigoori, 672-1297, and RIKEN (Institute of Physical and Chemical Research), Sendai 980-0845, Japan}
\author{K. Saito}
\affiliation{Department of Physics, The University of Tokyo, Tokyo 113-0033, Japan}
\author{A. Ohtomo}
\author{M. Kawasaki$^{b)}$}
\altaffiliation{Also at: Combinatorial Materials 
Exploration and Technology, NIMS, Tsukuba, 305-0047, Japan}
\affiliation{Institute for Materials Research, Tohoku University, Sendai, 
980-8577}
\author{R. T. Senger}
\affiliation{Department of Physics, Bilkent University, 06800 Ankara, Turkey}
\author{K. K. Bajaj}
\affiliation{Department of Physics, Emory University, Atlanta, Georgia{\nobreakspace}30322}

\date{\today}

\begin{abstract}
Localization dynamics of excitons was studied for ZnO/MgZnO and CdZnO/MgZnO quantum wells (QW). The experimental photoluminescence (PL) and absorption data were compared with the results of Monte Carlo simulation in which the excitonic hopping was modeled. The temperature-dependent PL linewidth and Stokes shift were found to be in a qualitatively reasonable agreement with the hopping model, with accounting for an additional inhomogeneous broadening for the case of linewidth. The density of localized states used in the simulation for the CdZnO QW was consistent with the absorption spectrum taken at 5~K.
\end{abstract}
\maketitle

\newpage

Recently, there have been extensive studies on optical properties 
of semiconductor quantum structures due to their size-dependent 
optical responses~\cite{efros1}. In particular, wide-gap II-VI oxides have attracted 
a great deal of attention owing to the fact that their excitons 
have large binding energies~\cite{lookp-type,nature_mat_tsukazaki}. The recent success of p-type doping~\cite{lookp-type,nature_mat_tsukazaki}
in ZnO has opened a door to realize practical application of 
blue and ultraviolet (UV) light-emitting diodes, which 
can be an alternative to those based on III-nitrides. Introduction 
of cadmium and/or magnesium into ZnO plays a key role in strain~\cite{makino17} 
and band engineering of these oxides~\cite{makino14,makino_sst}. In previous work, we have 
reported an ``anomalous'' S-shaped temperature dependence of the 
photoluminescence (PL) band peak that has been observed in ZnO/MgZnO and CdZnO/MgZnO multiple 
quantum wells (MQW)~\cite{makino12,makino13,makino_jap_2003}.

Here, in addition to the S-shaped dependence, we report the observation
of a W-shaped temperature dependence of the PL linewidth~\cite{kazl1}. For 
quantitative discussion, we model these anomalous temperature 
behaviors of the emission band and compare the results with the 
experimental data. We use a Monte Carlo simulation which is based 
on the phenomenological treatment of the excitonic kinetics~\cite{kazl1}. 
The comparison between simulation results and experimental data 
enables us to deduce the energy scale of distribution of the 
localized excitonic states.

The samples investigated here were ZnO and CdZnO QWs, whose growth and characteristics
have been described elsewhere. Both of these QWs were grown with laser molecular-beam epitaxy on ScAlMgO$_4$ substrates, while the CdZnO QW includes a ZnO buffer layer. The material of our barrier layers is 5-nm-thick Mg$_{x}$Zn$_{1-x}$O. The PL was studied in the 
temperature range from 5 to 300~K. The continuous-wave He-Cd laser 
(emitting wavelength being 325~nm) was used~\cite{makino8}.

The luminescence spectra taken for the ZnO/Mg$_{0.27}$Zn$_{0.73}$O QW (the well width being 1.8~nm) exhibit a double band~\cite{makino12,makino24} at intermediate temperatures of 95 to 175~K. Figure~1 shows a typical near-band edge PL spectrum measured at 120~K. These two emission bands correspond to radiative recombination of localized excitons (LX$_1$ and LX$_2$). The temperature dependence of the peak (see also Fig.~2 of Ref.~\onlinecite{makino12}) exhibits a well-known S-shaped variation~\cite{makino12}, it slightly red-shifts in the
range from 5 to 50~K, then blue-shifts in the range up to $\approx$100~K, splitts into a doubled peak at temperatures of 100 to $\approx $200~K, and then red-shifts again at higher temperatures. The initial
red-shift reflects the ability of localized excitons to reach 
lower-energy sites via thermally activated hopping. The blue 
shift and the splitting of peaks that are observed with further increase in temperature may be attributed to the population of the localized exciton states 
with increasingly higher energy, whereas the subsequent red-shift 
is due to the typical band gap shrinkage with temperature.

\begin{figure}[h]
\includegraphics[width=0.75\linewidth]{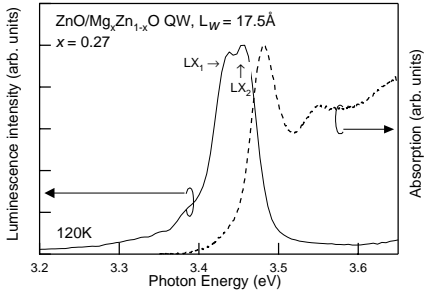}
\caption{The excitonic absorption and PL spectra in a
ZnO/Mg$_{0.27}$Zn$_{0.73}$O MQW (well width being 1.8~nm) taken at 120~K.}
\end{figure}
The observed features of hopping of localized excitons in our QWs are 
supported by the line width and Stokes shift measurements. The points in Fig.~2(a) and 2(b) show temperature variations of the full width at half maximum (FWHM) and Stokes shift of the PL band. Here, we notice a W-shaped (decrease-increase-decrease-increase) 
temperature dependence~\cite{kazl1} of the line width with a characteristic plateau at temperatures of 100 to 
175~K. Its abrupt increase with temperature (5--100~K) was related to thermalization of the excitons over localized 
states, as opposed to the situation at lower temperatures where the excitons do 
not attain their equilibrium distribution. This is a typical manifestation 
of hopping of localized excitons. On the other hand, the temperature variation of Stokes shift is more complicated due to the peak splitting at the intermediate temperatures. We show that these experimental results qualitatively 
agree with Monte-Carlo simulations explained later.

The observed temperature dependences were
simulated by a two-dimensional (2D) Monte Carlo algorithm. This algorithm
adopts the Miller-Abrahams rate for phonon-assisted exciton tunneling between
initial and final states (\textit{i} and \textit{j}) with the energies of \ensuremath{\epsilon}$_{{i}}$ and \ensuremath{\epsilon}$_{{j}}$ , respectively~\cite{kazl1}:
\begin{equation}
{\nu}_{{ij}} = {\nu}_{0} \exp \left[- \frac{2{r}_{{ij}}}{\alpha} - \frac{{\epsilon}_{{i}}-{\epsilon}_{{j}}+{|}{\epsilon}_{{i}}-{\epsilon}_{{j}}{|}}{2{kT}}\right]
\label{hoppingrate}
\end{equation}
Here ${r}_{{ij}}$ is the distance between the localization sites, ${\alpha} $
is the decay length of the exciton wave function, and ${\nu}_{0}$ is 
the attempt-escape frequency. Hopping was simulated over a randomly 
generated set of localized states with the sheet density of \textit{N}. Density 
of states (DOS) of the localization energies was assumed to be 
in accordance with a Gaussian distribution,

\begin{equation}
{g}({\epsilon})=( {N}^{2}/2{\pi}{\sigma}^{2} )^{2} \exp \left(- 
({\epsilon}-\textit{E}_{0})^{2}/2{\sigma}^{2} \right),
\label{gauss}
\end{equation}
with the peak positioned at the mean excitonic energy \textit{E}$_{0}$ 
and the dispersion parameter (the energy scale of the band potential 
profile fluctuation) ${\sigma}$. The reference energy was ${\epsilon}={E}_{0}$, which 
is the center of the distribution. All the energies in our simulation 
were below this value; i.e., we employed a ``half'' distribution~\cite{dal-don1}. In 
the case of the Gaussian DOS, the energies were also assumed 
to be uncorrelated. For each generated exciton, the hopping process 
terminates by recombination with the probability ${\tau}_{0}^{-1}$ and 
the energy of the localized state, where the recombination has 
taken place, is scored to the emission spectrum.

\begin{figure}[htb]
\includegraphics[width=0.75\linewidth]{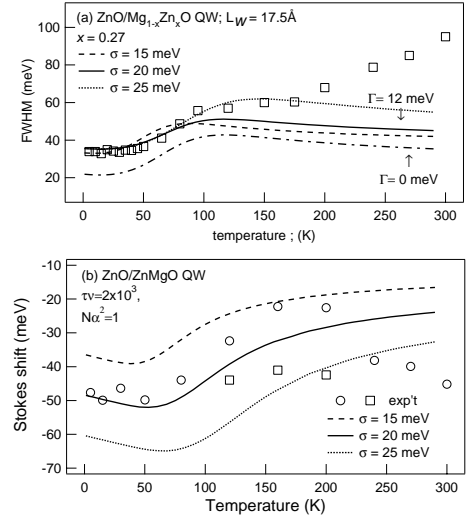}
\caption{(a) Evolution of the FWHM of the PL band with temperature 
in a ZnO MQW. Open squares show experimental data; dashed, dotted, 
and solid lines depict results of the Monte Carlo simulation of the exciton 
hopping for different scales of random potential fluctuations \ensuremath{\sigma} (indicated) 
with the inhomogeneous broadening \ensuremath{\Gamma} taken into account; the dash-dotted 
line represents results for \ensuremath{\sigma}=20 meV and \ensuremath{\Gamma}=0. (b) Simulated and experimental temperature dependences of the Stokes shifts of the PL
positions with respect to the excitonic transition energies.}
\end{figure}
\begin{figure}[htb]
\includegraphics[width=0.75\linewidth]{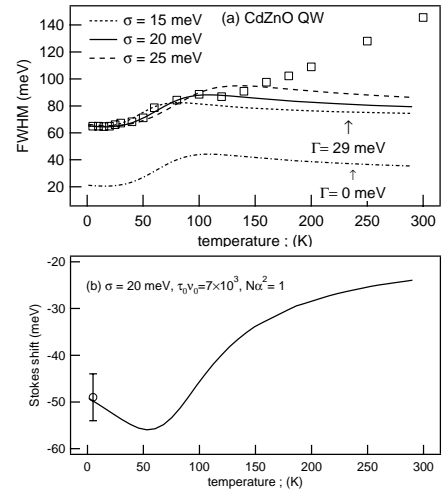}
\caption{Same as Fig.~2 except the sample is now a Cd$_{0.04}$Zn$_{0.96}$O/Mg$_{0.12}$Zn$_{0.88}$O QW (well width
being 1.8~nm).}
\end{figure}

The temperature variation of the data
is basically determined by the spatial (the product \textit{N}${\alpha}^{2}$) 
and temporal (the product ${\tau}_{0}{\nu}_{0}$) parameters 
of the hopping processes. For example, the kink in the temperature dependence
of the FWHM is 
related to the energy of the potential fluctuations, ${\sigma}$, i.e. $\sigma = 2 k_{\rm B} T_{\rm kink} $ (Ref.~\onlinecite{zimmermann1,kazl1}). A slightly decreasing behavior just above the saturation temperature 
(\textit{T}$_{kink}$) indicates the formation of a thermalized exciton 
distribution.

It has been difficult to perfectly reproduce both of our FWHM and energy shifting
bahaviors with the identical set of simulation parameters. Nevertheless,
if we adopt the values of \textit{N}${\alpha}^{2} {=}1$,
${\tau}_{0} {\nu}_{0} =2 {\times}10^{3}$, and ${\sigma}{=}20$~meV, the calculated result approximately reproduces the 22-meV amplitude of the variation of FWHM
curve as shown by a solid curve in Fig.~2(a). The dashed and dotted lines in Fig.~2(a) demonstrate the sensitivity of the simulated results with respect to the hopping energy scale ${\sigma}$. Increased $\sigma$ matches better with the
experiment, but, as explained later, we will meet serious difficulty in
accommodating the experimental Stokes shifts.
The quantitative agreement with the experimental 
data requires an additional inhomogeneous broadening ($\Gamma=12$~meV) to be introduced~\cite{kazl1}. In other words, our Monte Carlo simulation analysis reveals that 
two kinds of inhomogeneities (${\sigma}$ and ${\Gamma}$) are inherent to our QW.
The dash-dotted line shows the simulated dependence without inhomogeneous broadening ($\Gamma$).
The further increase of the linewidth above 175~K is attributed to 
the participation of the longitudinal optical phonons in the radiative 
transition and to the influence of delocalized excitons that 
are not accounted for in the model adopted here.

The temperature behavior of the Stokes shift deduced from the 
peak position of the simulated spectra is depicted in Fig.~2(b). The following feature was reproduced 
in the simulated results; the crossover temperature for the 
energy minimum in the S-shaped dependence (50--60~K). On the other hand,
the initial red-shifting behavior at low temperatures is somewhat exaggerated.
We can accomodate this disagreement if the value of $\tau_0 \nu_0$ is reduced.
But, in such a case, the agreement for the FWHM data becomes worse. Moreover, the simulation results do not reproduce the peak splitting phenomenon observed at the intermediate temperatures, the curve of which approximately runs in the
half way of the dual peaks (open circles and squares). The more refinement of the model should be necessary to describe this splitting. Nevertheless, it is thought that the simulated Stokes shift energies correspond to the average energy of the splitted peaks observed in the experiments. At temperatures higher than 230~K, the experimental data of the
Stokes shift are not in good agreement with those of simulation. This is probably because the spectral weight moves to the direction of the emission peak of longitudinal-optical phonon replicated recombination at the elevated temperatures as has been pointed out for ZnO bulk crystals and thin films~\cite{w_shan1,makino_jap_2005_2}.

For comparison, the simulated results for the Cd$_{0.04}$Zn$_{0.96}$O/Mg$_{0.12}$Zn$_{0.88}$O QW (the well width being 1.8~nm) were shown in Fig.~3(a) and 3(b). The solid line in Fig.~3(a) represents the best fit obtained for the following values of the parameters: \textit{N}${\alpha}^{2} {=}1$, ${\tau}_{0} {\nu}_{0} =7 {\times}10^{3}$, $\Gamma=29$~meV, and ${\sigma}{=}20$~meV. Unfortunately, the experimental Stokes shift datum
was obtained solely at 5~K. This is in a good agreement with the result of simulation as shown with the open circle in Fig.~3(b).

\begin{figure}[h]
\includegraphics[width=0.75\linewidth]{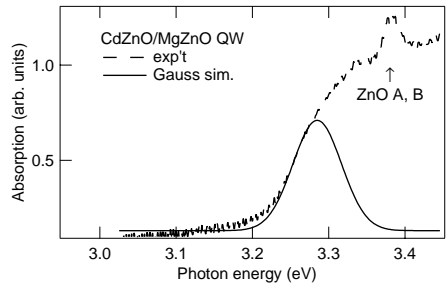}
\caption{Optical absorption spectrum (dashed line) of a CdZnO MQW taken at 5~K.
A solid line shows the density of band-tail states used in the simulation
with the inhomogeneous broadening (\ensuremath{\Gamma}) accounted for. The
most prominent peaks (``ZnO A, B'') come from the ZnO buffer layer.}
\end{figure}
Finally, Fig.~4 compares the 5~K absorption spectrum (a dashed line) with the actual density of localized states used in the simulation (a solid line),

\begin{equation}
D({E}){\propto} \exp[\frac{-(E-E_{0})^{2}}{2({\sigma}^{2}+{\Gamma}^{2})}],
\label{eq:BEDL}
\end{equation}
where the inhomogeneous broadening is taken into account by introduction of $\Gamma$ (here 29~meV). These densities of band-tail states are centered at the exciton transition energy (3.285~eV) obtained by adding the \textit{simulated} Stokes shift [Fig.~3(b)] with the \textit{experimental} PL energy (see also Fig.~2 of Ref.~\onlinecite{makino13}). In addition, the long-wavelength wing of $D(E)$
is seen to be in a fairly good coincidence with the low-energy tail of the corresponding absorption spectrum. This is thought to be an additional proof on the
validity of the exciton hopping model.

The value of $\Gamma$ for the CdZnO QW (29~meV) is significantly larger than that for the ZnO QW (12~meV),
which is not surprising because of the more severe inhomogeneity in the former sample;
both the well and barrier layers are comprised of alloyed materials.
Accordingly, $\tau_0 \nu_0$ for the CdZnO QW is also larger. It is likely that
the effective lifetime (related to $\tau_0$) is longer because it is known that
the non-radiative decay path becomes reduced for more deeply localized
excitons~\cite{makino13}.

Here we try to evaluate precise Cd concentration at irradiated region of the sample
from a variational calculation~\cite{senger1} where the electron--phonon interaction is properly taken
into account. In general, the precise determination is difficult for Cd doped samples because of the reevaporation during the growth or to the
large inhomogeneity. If the concentration is set to 4.3\%, the transition energy is calculated to
be 3.288 eV, which is very close to the exciton transition energy at 5~K.

In conclusion, we have studied evolution of the PL maximum and FWHM
of ZnO and CdZnO QWs with temperature by the Monte Carlo simulation. By making
efforts to reproduce both of our experimental FWHM and energy shift data with the
same set of parameters consistently, characteristic energy scale of the distribution
of the localized states for the ZnO QW was evaluated to be \ensuremath{\sigma} $\approx$20~meV. The 
simulation revealed two kind of inhomogeneities (\ensuremath{\sigma} and
\ensuremath{\Gamma}) are inherent to our ZnO-related QWs. The values of
$\Gamma$ and $\tau_0 \nu_0$ for the CdZnO QW are significantly larger
than those for the ZnO QW.
The density of localized states employed in the simulation turned out to
be in a good agreement with the long-wavelength region of the absorption
spectrum.

\begin{center}
Acknowledgements
\end{center}

The authors are thankful to Y. Segawa, N. T. Tuan, C. H. Chia, N. Shima, and Y. Takagi for helpful discussions and for providing the experimental data. Thanks are also due to H. Koinuma for encouraging our work. This work was partially supported by MEXT Grant of Creative Scientific Research 14GS0204, the Asahi Glass Foundation, and the inter-university cooperative program of the IMR, Japan.

\newpage 

\begin{thebibliography}{10}

\bibitem{efros1}
A.~L. Efros, M. Rosen, M. Kuno, M. Nirmal, D.~J. Norris, and M. Bawendi, Phys.
  Rev. B {\bf 54},  4843  (1996).

\bibitem{lookp-type}
D.~C. Look, D.~C. Reynolds, C.~W. Litton, R.~L. Jones, D.~B. Eason, and G.
  Cantwell, Appl. Phys. Lett. {\bf 81},  1830  (2002).

\bibitem{nature_mat_tsukazaki}
A. Tsukazaki, A. Ohtomo, T. Onuma, M. Ohtani, T. Makino, M. Sumiya, K. Ohtani,
  S.~F. Chichibu, S. Fuke, Y. Segawa, H. Ohno, H. Koinuma, and M. Kawasaki,
  Nat. Mater. {\bf 4},  42  (2005).

\bibitem{makino17}
T. Makino, Y. Segawa, A. Ohtomo, K. Tamura, T. Yasuda, M. Kawasaki, and H.
  Koinuma, Appl. Phys. Lett. {\bf 79},  1282  (2001).

\bibitem{makino14}
T. Makino, Y. Segawa, M. Kawasaki, A. Ohtomo, R. Shiroki, K. Tamura, T. Yasuda,
  and H. Koinuma, Appl. Phys. Lett. {\bf 78},  1237  (2001).

\bibitem{makino_sst}
T. Makino, Y. Segawa, M. Kawasaki, and H. Koinuma, Semicond. Sci. \& Technol.
  {\bf 20},  S78  (2005).

\bibitem{makino13}
T. Makino, N.~T. Tuan, Y. Segawa, C.~H. Chia, M. Kawasaki, A. Ohtomo, K.
  Tamura, and H. Koinuma, Appl. Phys. Lett. {\bf 77},  1632  (2000).

\bibitem{makino12}
T. Makino, N.~T. Tuan, H.~D. Sun, C.~H. Chia, Y. Segawa, M. Kawasaki, A.
  Ohtomo, K. Tamura, M. Baba, H. Akiyama, T. Suemoto, S. Saito, T. Tomita, and
  H. Koinuma, Appl. Phys. Lett. {\bf 78},  1979  (2001).

\bibitem{makino_jap_2003}
T. Makino, C.~H. Chia, K. Tamura, A. Ohtomo, M. Kawasaki, H. Koinuma, and Y.
  Segawa, J. Appl. Phys. {\bf 93},  5929  (2003).

\bibitem{kazl1}
K. Kazlauskas, G. Tamulaitis, A. Zukauskas, M.~A. Khan, J.~W. Yang, J. Zhang,
  G. Simin, M.~S. Shur, and R. Gaska, Appl. Phys. Lett. {\bf 83},  3722
  (2003).

\bibitem{makino8}
T. Makino, C.~H. Chia, N.~T. Tuan, Y. Segawa, M. Kawasaki, A. Ohtomo, K.
  Tamura, and H. Koinuma, Appl. Phys. Lett. {\bf 76},  3549  (2000).

\bibitem{makino24}
T. Makino, K. Tamura, C.~H. Chia, Y. Segawa, M. Kawasaki, A. Ohtomo, and H.
  Koinuma., Phys. Rev. B {\bf 66},  233305  (2002).

\bibitem{dal-don1}
B.~D. Don, K. Kohary, E. Tsitsishvili, H. Kalt, S.~D. Baranovskii, and P.
  Thomas, Phys. Rev. B {\bf 69},  045318  (2004).

\bibitem{zimmermann1}
R. Zimmermann, F. Grosse, and E. Runge, Pure Appl. Chem. {\bf 69},  1179
  (1997).

\bibitem{w_shan1}
W. Shan, W. Walukiewicz, J.~W. AgerIII, K.~M. Yu, H.~B. Yuan, H.~P. Xin, G.
  Cantwell, and J.~J. Song, Appl. Phys. Lett. {\bf 8},  191911  (2005).

\bibitem{makino_jap_2005_2}
T. Makino, Y. Segawa, S. Yoshida, A. Tsukazaki, A.~Ohtomo, M. Kawasaki, and H.
  Koinuma, to appear in the Nov. issue of J. Appl. Phys.  (2005).

\bibitem{senger1}
R.~T. Senger and K.~K. Bajaj, Phys. Rev. B {\bf 68},  205314  (2003).

\end{thebibliography}

\end{document}